# Diamond-Like C$_2$H Nanolayer, Diamane: Simulation of the Structure and Properties

L. A. Chernozatonskii[a], P. B. Sorokin[a,b], A. G. Kvashnin[b], and D. G. Kvashnin[b]

[a]*Emanuel Institute of Biochemical Physics, Russian Academy of Sciences, ul. Kosygina 4, Moscow, 119334 Russia*
*email: cherno@sky.chph.ras.ru*
[b]*Siberian Federal University, Krasnoyarsk, 660041 Russia*

We consider a new C$_2$H nanostructure based on bilayer graphene transformed under the covalent bond of hydrogen atoms adsorbed on its external surface, as well as compounds of carbon atoms located opposite each other in neighboring layers. They constitute a "film" of the $\langle 111 \rangle$ diamond with a thickness of less than 1 nm, which is called diamane. The energy characteristics and electron spectra of diamane, graphene, and diamond are calculated using the density functional theory and are compared with each other. The effective Young's moduli and destruction thresholds of diamane and graphene membranes are determined by the molecular dynamics method. It is shown that C$_2$H diamane is more stable than CH graphane, its dielectric "gap" is narrower than the band gap of bulk diamond (by 0.8 eV) and graphane (by 0.3 eV), and is harder and more brittle than the latter.

## INTRODUCTION

Carbon nanomaterials have been recently supplemented by structures based on graphene, which is a single-atom layer [1–4]. They attract attention owing not only to unique properties, but also to prospects of wide application in electronics, sensorics, materials science, and biomedical applications (see references in [1, 3]). It is also known that nanostructures consisting of two (bilayer graphene with the AB Bernal structure) and more graphene layers are obtained with properties different from graphene itself [2].

Recently, placing a graphene sample to a discharge hydrogen plasma, where a gas was decomposed into hydrogen ions, Elias et al. [1] obtained a graphene layer enriched in covalently bonded hydrogen atoms and, thus, transformed highly conducting graphene into an insulator. The possibility of such a rearrangement of the electronic properties was pointed out in works [2–5] on structures with chemically adsorbed hydrogen. Sofo et al. [6] predicted layered hydrocarbon CH, graphane, whose structure is graphene with periodically "attached" hydrogen atoms from both of its surfaces, so that each carbon atom is in a diamond like $sp^3$ hybridized state, see Fig. 1a. The successful experimental hydration of graphene was recently reported in [7, 8].

In this work, we consider a new layered structure C$_2$H based on bilayer graphene, where each carbon atom is $sp^3$ hybridized, as in graphene. The carbon atoms of one of two atomic sublattices of the graphene layer are covalently bonded with hydrogen atoms and the carbon atoms of the other sublattice that are located above the carbon atoms of the neighboring graphene are covalently attached to the latter atoms, see Fig. 1b. We call such a C$_2$H structure diamane by analogy with the hydrocarbon molecules and nanostructures with different hybridizations (e.g., alkenes and alkanes, graphene and graphane). We also consider a diamane isomer, diamane-II, obtained from the bilayer AA graphene (see Fig. 1c), which is also very stable.

We assume that diamane can be formed from bilayer graphene placed in a discharge hydrogen plasma. In this case, under corresponding conditions (the pressure and temperature of a hydrogen gas), hydrogen atoms are chemically adsorbed on both of its sides. Thus, a carbon atom of the "upper" graphene, which is not located over a carbon atom of the "lower" graphene and, therefore, is freer, after the attachment of a hydrogen atom to the carbon atom, is displaced from the plane owing to $sp^3$ hybridization. In this case, three carbon atoms surrounding it are displaced downward. A similar rearrangement occurs with the atoms of the "lower" graphene, which leads to the $sp^3$ hybridization of carbon atoms from neighboring layers located over each other. This leads to the formation of the diamane nucleus, see Fig. 1d.

We perform the optimization of the structure of diamanes and graphane. The molecular dynamics simulation of the punching of the membrane by a nanotip indicates that diamane has a larger effective stiffness coefficient than graphane (it is stronger), but is destroyed at lower applied loads.

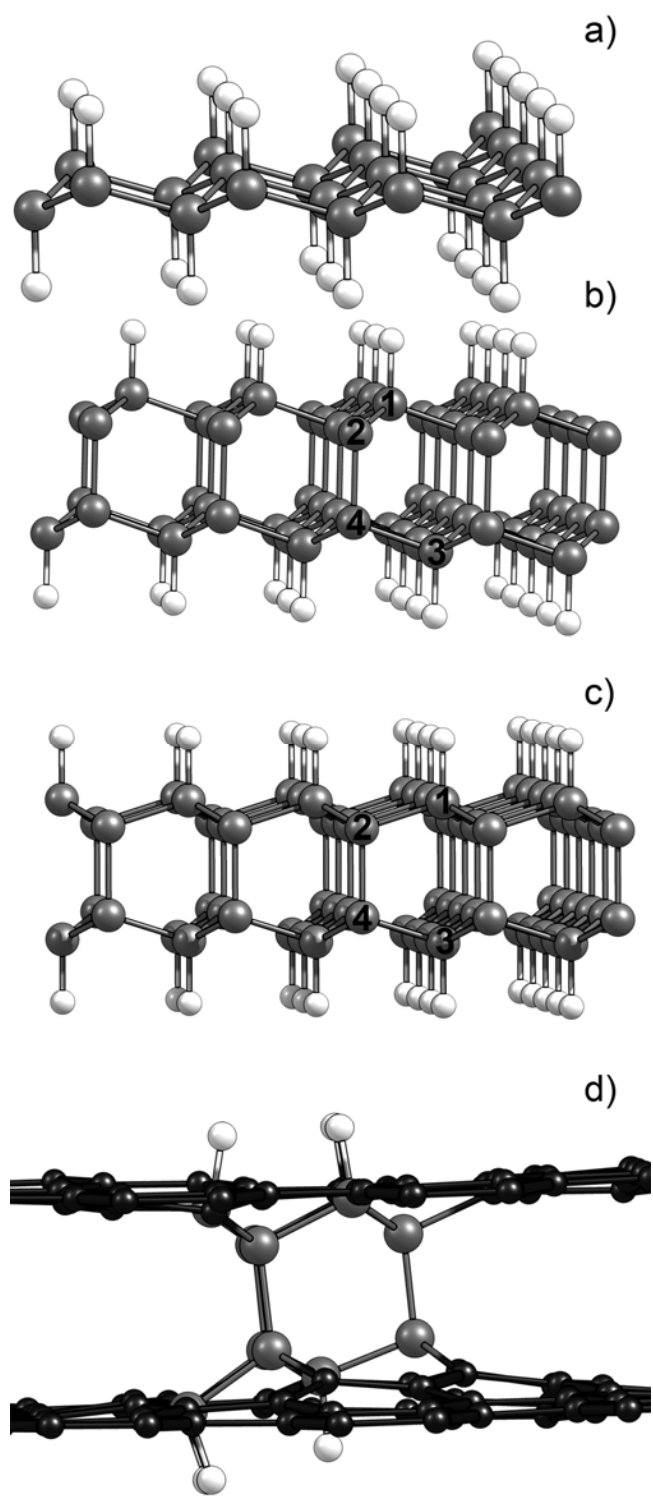

**Fig. 1.** Atomic structures of (a) graphane, (b) diamane, and (c) diamane-II and (d) the scheme of the formation of a diamane nucleus on an initial bilayer graphene: hydrogen atoms settle from two sides and initiate the "adhesion" of carbon atoms located over each other in neighboring layers.

## CALCULATION METHODS

The calculations were performed with the Vienna Ab Initio Simulation Package (VASP) [9–11]. This package for ab initio calculations is based on the Vanderbilt pseudopotential [12] and the expansion in the plane wave basis in the local density functional formal ism [13–15]. The optimization of geometry was per formed with 4 × 4 × 1 $k$ points in the Brillouin zone. The $k$ points were generated by the Monkhorst–Pack method [16]. The cutoff energy of plane waves was taken to be 286.7 eV.

The stiffness coefficient of nanomembranes of the structures under consideration was calculated with the molecular dynamics method (GULP package [17]).

The Brenner potential [18], which well describes carbon structures [19], was used in the calculations. The potential between the needle atoms and graphene was taken to be repulsive in order to avoid nonrealistic bonding between atoms.

## STRUCTURE AND ENERGY CHARACTERISTICS OF DIAMANE

We calculate the optimal configurations of diamane and diamane-II and compare their geometric parameters with the characteristics of graphane and diamond. They are presented in the table. The atoms are numbered according to the notation in Figs. 1b and 1c.

According to the table, the arrangement of atoms is very close to the structure of bulk diamond [20]. To estimate the stability of the structure under investigation, we calculate the energies of the formation of graphane and diamane using the following equation [6]

$$E_{form} = (E_{str} - nE_{graphene} - 2E_{H_2})/(n+2),$$

where $E_{str}$ is the total energy of a structure (graphene or diamane), $E_{graphene}$ is the carbon atom energy in graphene, $E_{H_2} = -3.38$ eV/atom is the energy of a hydrogen atom in a $H_2$ molecule, and $n$ is the number of carbon atoms in a unit cell of the structure.

Atomic geometries of graphane, diamanes, and diamond. The angle between the carbon and hydrogen atoms is given for graphane

|  | $d_{C1(3)-H}$, Å | $d_{C2-C4}$, Å | $d_{C1(3)-C2(4)}$, Å | $\angle(C1,C2,C4)$ |
|---|---|---|---|---|
| Graphane | 1.11 | 1.54 |  | 107.4° |
| Diamane | 1.10 | 1.52 | 1.53 | 107.9° |
| Diamane-II | 1.12 | 1.58 | 1.51 | 107.7° |
| Diamond |  | 1.54 |  | 109.5° |

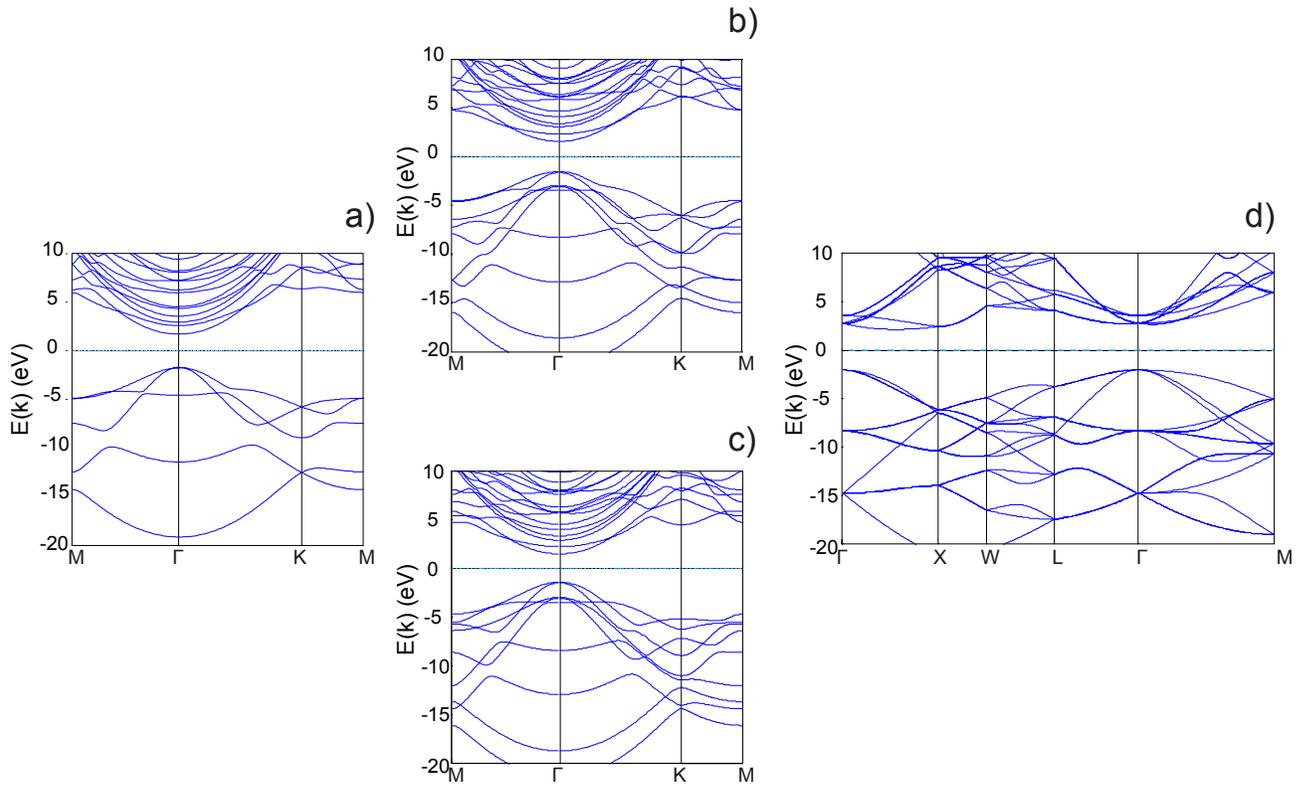

**Fig. 2.** Band structures of (a) graphane CH, (b) diamane, (c) diamane-II, and (d) diamond. The Fermi level is taken as zero and marked by the horizontal dashed line.

The calculated graphane formation energy, -0.11 eV/atom, is close to the value of –0.15 eV/atom obtained in [6]. Diamane is much more stable than graphane: its formation energy is –0.70 eV/atom. The formation energy of diamane-II is –0.69 eV/atom.

## ELECTRONIC CHARACTERISTICS OF DIAMANE IN COMPARISON WITH GRAPHANE AND DIAMOND

The calculated band structures of diamane, graphene, graphane, and diamond are shown in Fig. 2.

First, we determined the dielectric gaps $E_g$ for graphane (3.42 eV) and diamond (4.19 eV). It is necessary to take into account that the local electron density approximation used in this work underestimates the energy gap; for example, the experimental value for pure diamond is 5.45 eV [21]. It is seen that the character of the spectrum near the Γ point of both graphane CH (see Fig. 2a) and diamane $C_2H$ (see Fig. 2b) qualitatively coincides with the band structure of bulk diamond (see Fig. 2d). The band gaps for diamane and diamane-II are 3.12 and 2.94 eV, respectively. Thus, diamane is an insulator with a somewhat narrower gap than that of graphane.

## ELASTIC PROPERTIES OF DIAMANE

The elastic characteristics of carbon nanofilms were investigated in the framework of the simulation of the experimental scheme of the punching of the membrane by the tip of an atomic force microscope [22]. We also used this scheme to compare the elastic properties of diamane, graphene, and graphane. Since the elastic energy $E$ rather than the force $F$ is a measured quantity in theoretical simulation, the dependence of the former on the deflection depth (see Fig. 3) is approximated by the fourth order polynomial

$$E = b \cdot \left(\frac{\delta}{d}\right)^4 + c \cdot \left(\frac{\delta}{d}\right)^2,$$

from which the coefficient $b$ is calculated. Using the formula

$$b = \frac{E^{2D} q^3}{4d^2}$$

the stiffness coefficient $E^{2D}$ is calculated. Here, $d$ is the membrane radius and $q = 1/(1.05 - 0.15\nu - 0.16\nu^2)$ is the dimensionless constant, where $\nu = 0.2$ is the Poisson coefficient for diamond [21]. Since the layer thickness for diamane, as well as for graphene, is an uncertain quantity, we use $E^{2D}$ to estimate the stiffness of the structure.

We consider graphene, graphane, and diamane membranes with the same radius (30 Å) with fastened edges. Deflection was carried out with a step of 0.1 Å; the optimization of geometry was performed at each step by the annealing method with an initial temperature of 2000 K.

Diamane appeared to be stronger than graphane and graphene; it has a much larger stiffness coefficient $E^{2D}_{graphene} = 238$ N/m, $E^{2D}_{graphane} = 449$ N/m и $E^{2D}_{diamane} = 715$ N/m), but is more brittle: it is destroyed at lower applied loads; the critical deflections of the critical structure under investigation are $\delta^{critical}_{graphane} = 8.9$ Å and $\delta^{critical}_{diamane} = 7.6$ Å.

Diamanes $C_2H$ with $n > 2$ can also be considered, but their formation is much more difficult owing to more stringent conditions on the synthesis parameters (pressure/temperature of a hydrogen gas). The properties of such structures (geometric structure, electronic properties) are expected to tend to the proper ties of the diamond crystal.

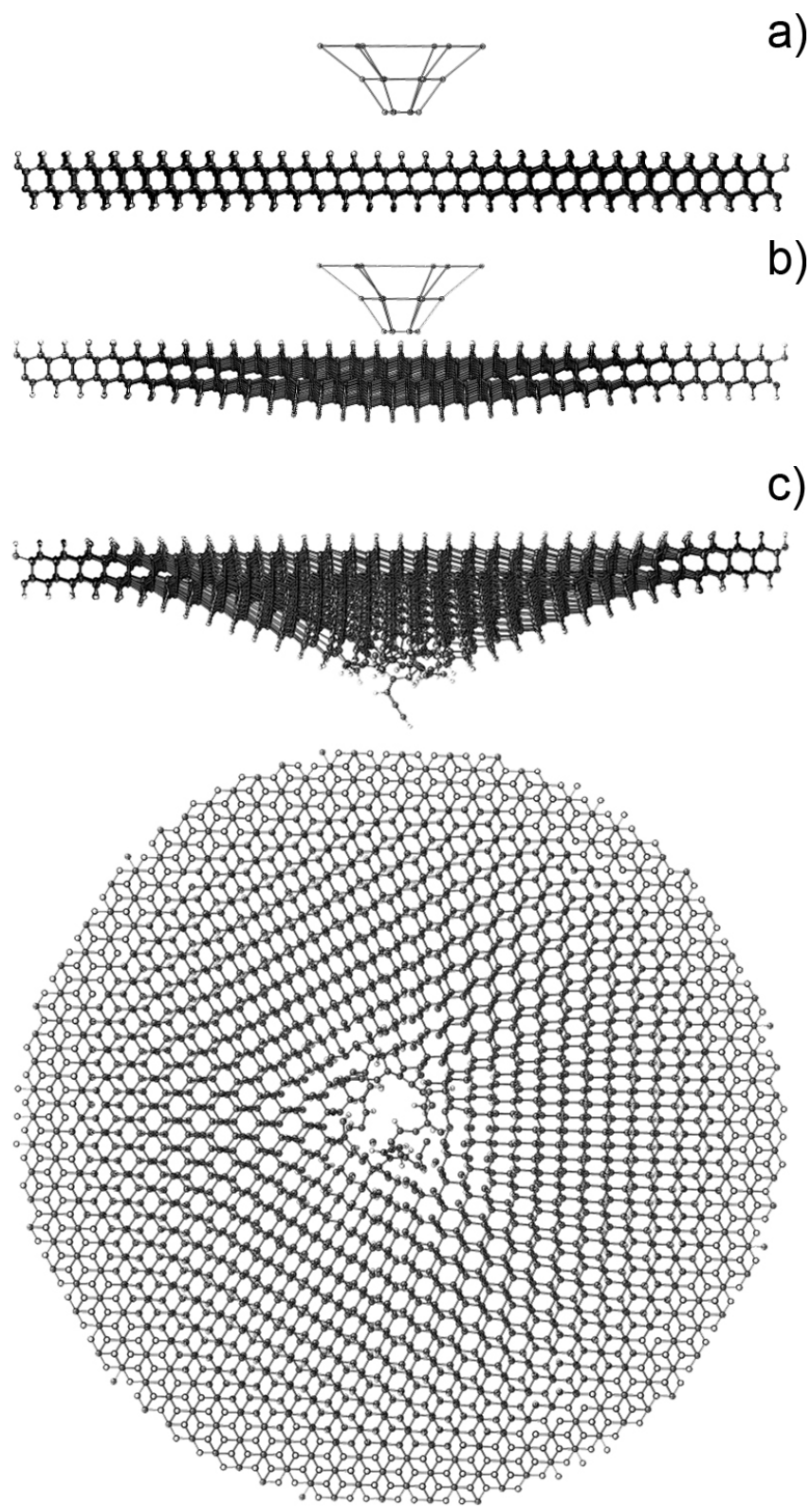

**Fig. 3.** Molecular dynamics simulation of the deformation of the diamane membrane: (a) initial, undeformed structure; (b) elastic deformation; and (c) membrane break (atoms simulating the tip of an atomic force microscope are connected for convenient perception).

CONCLUSIONS

Thus, we have shown that diamane $C_2H$, which is a diamond nanofilm, obtained from bilayer graphene after its hydrogenization should be significantly different from graphene and its completely hydrogenised analog, graphane, in electric and mechanical properties. Similar to graphene–graphane structures, where graphane dielectric nanostrips [5] (or simply "diamond lines" consisting of adsorbed hydrogen atoms [4]) can form nanowaveguides on graphene, the formation of similar structures such as superlattices or integral nanowaveguides is expected on the basis of bilayer graphane ruled by diamane nanostrips. Moreover, diamane films will be interesting in applications as ultrathin dielectric nanolayers.

Note an interesting feature (which was missed by the authors of the first calculation of graphane [6]): the crystal structures of graphane (see Fig. 1b) and diamane (see Fig. 1c) do not contain an inversion center; this indicates the presence of piezoelectricity in them. This effect can be very interesting for the use of diamane in electronics and electromechanical microsystems, and the presence of a direct energy gap in graphane and diamane can be used in nanophotonics.

We are grateful to I.V. Stankevich and A.O. Litinskii for discussions. This work was supported by the Russian Foundation for Basic Research (project nos. 08-02-01096 and 08-03-00420) and the Russian Academy of Sciences (program no. 27).